\documentclass[
    ,final            
  ]
  {aipproc}

\usepackage{amsmath}
\layoutstyle{6x9}

\newcommand{\be}{\begin{equation}}
\newcommand{\ee}{\end{equation}}
\newcommand{\bea}{\begin{eqnarray}}
\newcommand{\eea}{\end{eqnarray}}
\newcommand{\bi}{\begin{itemize}}
\newcommand{\ei}{\end{itemize}}

\newcommand{\bc}{\begin{center}}
\newcommand{\ec}{\end{center}}

\newcommand{\Pslash}{p \kern -2mm /}
\newcommand{\Btob}{B\rightarrow b}

\begin{document}

\vspace*{-20mm}
\begin{flushright}
TKYNT-11-04
\end{flushright}

\title{Hyperon vector coupling $f_1(0)$ \\from 2+1 flavor lattice QCD}


\author{Shoichi Sasaki}{
  address={Department of Physics, The University of Tokyo, Hongo 7-3-1, 113-0033, Japan}
}

\begin{abstract}
 We present results for the hyperon vector form factor $f_1$ 
 for $\Xi^0 \rightarrow \Sigma^+ l\bar{\nu}$ and $\Sigma^- \rightarrow n l\bar{\nu}$ semileptonic 
 decays from dynamical lattice QCD with domain-wall quarks. Simulations are performed on
 the 2+1 flavor gauge configurations generated by the RBC and UKQCD Collaborations
 with a lattice cutoff of $a^{-1} = 1.7$ GeV. Our preliminary results, which are
 calculated at the lightest sea quark mass (pion mass down to approximately 330 MeV),
 show that a sign of the second-order correction of SU(3) breaking
 on hyperon vector coupling $f_1(0)$ is likely negative.
\end{abstract}

\maketitle


\section{Introduction}

\indent\indent
The matrix element for hyperon beta decays $B\rightarrow bl\bar{\nu}$
is composed of the vector and axial-vector transitions, $\langle b(p') | V_{\alpha}(x)+A_{\alpha}(x) | B(p) \rangle$,
which is described by six form factors: the vector ($f_1$), weak-magnetism $(f_2)$, and induced scalar $(f_3)$ form factors
for the vector current, and the axial-vector $(g_1)$, weak electricity $(g_2)$, and induced pseudo-scalar $(g_3)$ form factors
for the axial current. The experimental rate of the hyperon beta decay, $B\rightarrow bl\bar{\nu}$, is given by
\be
\Gamma=\frac{G_F^2}{60\pi^3}(M_B-M_b)^5(1-3\delta)
|V_{us}|^2|f^{\Btob}_1(0)|^2\left[1+3
\left|\frac{g^{\Btob}_1(0)}{f^{\Btob}_1(0)}\right|^2
+ \cdot\cdot\cdot
\right],
\ee
where $G_F$ and $V_{us}$ denote the Fermi constant and an element of the Cabibbo-Kobayashi-Maskawa (CKM)
mixing matrix respectively~\cite{Cabibbo:2003cu}. 
Here, 
$M_B$ ($M_b$) denotes the rest mass of the initial (final) state. 
The ellipsis can be expressed in terms of a power series in the small parameter $\delta=(M_B-M_b)/(M_B+M_b)$, which is regarded as a size of flavor SU(3) breaking~\cite{Gaillard:1984ny}. 
The first linear term in $\delta$, which should be given by 
$-4\delta [g_2(0)g_1(0)/f_1(0)^2]_{\Btob}$, 
is safely ignored as small as ${\cal O}(\delta^2)$ since the nonzero value of 
the second-class form factor $g_2$~\cite{Weinberg:1958ut} should be induced 
at first order of the $\delta$ expansion~\cite{Gaillard:1984ny}. 
The absolute value of $g_1(0)/f_1(0)$ can be determined
by measured asymmetries such as electron-neutrino 
correlation~\cite{{Cabibbo:2003cu},{Gaillard:1984ny}}.
Therefore a theoretical estimate of vector coupling $f_1(0)$ is primarily required for the precise 
determination of $|V_{us}|$.

The value of $f_1(0)$ should 
be equal to the SU(3) Clebsch-Gordan coefficients up to the second 
order in SU(3) breaking, thanks to the Ademollo-Gatto 
theorem~\cite{Ademollo:1964sr}. 
As the mass splittings among octet baryons are typically 
of the order of 10-15\%, an expected size of the second-order 
corrections is a few percent level. However, either the size, or the sign of 
their corrections are somewhat controversial among various 
theoretical studies at present as summarized in Table~\ref{Tab:Th_estimate_f_1}. 
A model independent evaluation of SU(3)-breaking corrections 
is highly demanded. Although recent quenched lattice studies suggest
that the second-order correction on $f_1(0)$ is likely negative~\cite{{Guadagnoli:2006gj},{Sasaki:2008ha}}, 
we need further confirmation from (2+1)-flavor dynamical lattice QCD near the
physical point.

\begin{table}[tbp]
\begin{tabular}{ l  |  l l l l   c }
\hline
Type of result
&$\Lambda\rightarrow p$ & $\Sigma^- \rightarrow n$ & $\Xi^- \rightarrow \Lambda$ & $\Xi^0
\rightarrow\Sigma^+$ & Reference\\
\hline
Bag model & 0.97 & 0.97 & 0.97 & 0.97 & \cite{Donoghue:1981uk}\\
Quark model & 0.987 & 0.987 & 0.987 & 0.987 & \cite{Donoghue:1986th}\\
Quark model &0.976&0.975&0.976&0.976 & \cite{Schlumpf:1994fb}\\
$1/N_c$ expansion  &1.02(2)&1.04(2)&1.10(4)&1.12(5) &\cite{Flores-Mendieta:1998ii}\\
Full ${\cal O}(p^4)$ HBChPT &1.027&1.041&1.043&1.009&\cite{Villadoro:2006nj}\\
Full ${\cal O}(p^4)$ + partial ${\cal O}(p^5)$ HBChPT & 1.066(32) & 1.064(6) & 1.053(22) & 1.044(26) &
\cite{Lacour:2007wm}\\
Full ${\cal O}(p^4)$ IRChPT &0.943(21) &1.028(02)& 0.989(17) & 0.944(16) & \cite{Geng:2009ik}\\
Full ${\cal O}(p^4)$ IRChPT + Decuplet 
&1.001(13) &1.087(42)& 1.040(28) & 1.017(22) & \cite{Geng:2009ik}\\
\hline
Quenched lattice QCD & N/A & 0.988(29) & N/A & 0.987(19)&\cite{{Guadagnoli:2006gj},{Sasaki:2008ha}}\cr
\hline
\end{tabular}
\caption{Theoretical uncertainties of 
$\tilde{f}_1=|f_1/f_1^{\rm SU(3)}|$ for various hyperon beta-decays.
HBChPT and IRChPT stand for heavy baryon chiral perturbation theory and
the infrared version of baryon chiral perturbation theory.
\label{Tab:Th_estimate_f_1}
}
\end{table}

\section{Numerical results}

\indent\indent
In this study, we use the RBC-UKQCD joint (2+1)-flavor dynamical 
DWF coarse ensembles on a $24^3\times 64$ lattice~\cite{Allton:2008pn}, 
which are generated with the Iwasaki gauge action at $\beta=2.13$.
For the domain wall fermions  with the domain-wall height of $M_5=1.8$, 
the number of sites in the fifth dimension is 16, which gives
a residual mass of $am_{\rm res}\approx 0.003$.
Each ensemble of configurations uses the same dynamical strange quark
mass, $am_{s}=0.04$. The inverse of lattice spacing is $a^{-1}=1.73(3)$ ($a$=0.114(2) fm), 
which is determined from the $\Omega^{-}$ baryon mass~\cite{Allton:2008pn}. 
We have already published our findings in nucleon structure from the same ensembles 
in three publications, Refs.~\cite{{Yamazaki:2008py},{Yamazaki:2009zq},{Aoki:2010xg}}.

In this study, we calculate the vector coupling $f_1(0)$ for two different hyperon beta-decays, 
$\Xi^0\rightarrow \Sigma^+ l\bar{\nu}$ and $\Sigma^-\rightarrow n l\bar{\nu}$, where
$f^{\Xi\rightarrow \Sigma}_1(0)=+1$ and $f^{\Sigma\rightarrow n}_1(0)=-1$ in the exact SU(3) 
limit. We will present our results for $am_{ud}=0.005$, which corresponds to about
330 MeV pion mass. We use 4780 trajectories (the range from 940 to 5720 in molecular-dynamics time)
separated by 20 trajectories. The total number of configurations is actually 240.
We make two measurements on each configuration using two locations
of the source time slice, $t_{\rm src}=0$ and 32.
Details of our calculation of the quark propagators are described in Ref~\cite{Yamazaki:2009zq}.

For convenience in numerical calculations, instead of the vector form factor $f_1(q^2)$, 
we consider the so-called scalar form factor
\be
f^{B\rightarrow b}_S(q^2)=f^{B\rightarrow b}_1(q^2)+\frac{q^2}{M_B^2-M_b^2}f^{B\rightarrow b}_3(q^2),
\ee
where $f_3$ represents the second-class form factor, which are identically zero in the exact SU(3) limit~\cite{Weinberg:1958ut}. 
The value of $f_S(q^2)$ at $q_{\rm max}^2=-(M_B-M_b)^2$ can be precisely 
evaluated by the double ratio method proposed in Ref.~\cite{Guadagnoli:2006gj},
where all relevant three-point functions are determined at zero three-momentum transfer $|{\bf q}|=0$.

Here we note that the absolute value of 
the renormalized $f_S(q_{\rm max}^2)$ is exactly unity in the flavor SU(3) symmetric limit, where $f_S(q^2_{\rm max})$ becomes $f_1(0)$, for the hyperon decays considered here. Thus, the
deviation from unity in $f_S(q^2_{\rm max})$ is attributed to three types of the SU(3) breaking effect:
(1) the recoil correction ($q^2_{\rm max}\neq 0$) stemming from the mass difference of $B$ and $b$ states,
(2) the presence of the second-class form factor $f_3(q^2)$, and 
(3) the deviation from unity in the renormalized $f_1(0)$. Taking the limit of zero four-momentum transfer
of $f_S(q^2)$ can separate the third effect from the others, since the scalar form factor at $q^2=0$, $f_S(0)$,
is identical to $f_1(0)$. Indeed, our main target is to measure the third one.

\begin{figure}
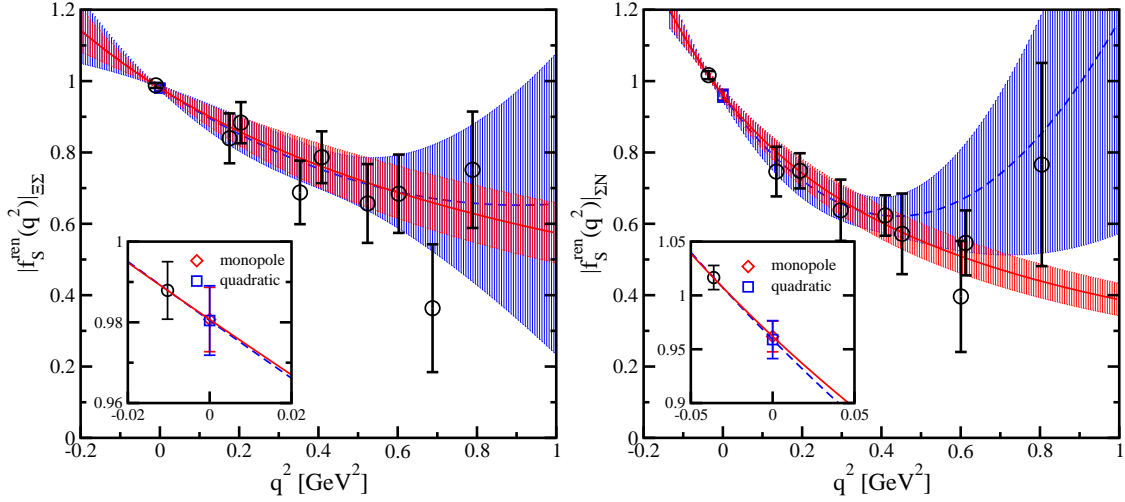

\includegraphics[width=.5\textwidth,clip]{q2_extra_F0_xi2sg_ml0005.eps}
\includegraphics[width=.5\textwidth,clip]{q2_extra_F0_sg2nu_ml0005.eps}
\caption{Interpolation of $|f_S(q^2)|$ to $q^2=0$
for $\Xi^{0}\rightarrow \Sigma^{+}$ (left) and $\Sigma^{-}\rightarrow n$ (right)
at $am_{ud}=0.005$.
}
\label{Fig:scalarFF}
\end{figure}

The scalar form factor $f_S(q^2)$ at $q^2>0$~\footnote{We note that
$q^2$ quoted here is defined in the Euclidean metric convention.} is calculable with non-zero three-momentum transfer
($|{\bf q}|\neq 0$)~\cite{Sasaki:2008ha}. We can make the $q^2$ interpolation of $f_S(q^2)$ to $q^2=0$
together with the precisely measured value of $f_S(q^2)$ at $q^2=q^2_{\rm max}<0$. In Fig.~\ref{Fig:scalarFF}, we plot
the absolute value of 
the renormalized $f_S(q^2)$ as a function of $q^2$ for $\Xi^{0}\rightarrow \Sigma^{+}$ (left) 
and $\Sigma^{-}\rightarrow n$ (right) at $am_{ud}=0.005$.
Open circles are $f_S(q^2)$ at the simulated $q^2$. The solid (dashed)
curve is the fitting result by using the monopole (quadratic) interpolation
form~\cite{Sasaki:2008ha}, while the open diamond (square) represents the interpolated value
to $q^2=0$. As shown in Fig.~\ref{Fig:scalarFF}, two determinations to
evaluate $f_S(0)=f_1(0)$ from measured points are indeed consistent
with each other. 

We finally quote the values obtained from the monopole fit
as our final values. The values of the renormalized $f_1(0)$ divided by the SU(3) symmetric
value at $m_\pi=330$ MeV are obtained as
\bea
[f_1(0)/f_1^{\rm SU(3)}]_{\Xi\rightarrow\Sigma} &=& 0.981(8), \cr
[f_1(0)/f_1^{\rm SU(3)}]_{\Sigma\rightarrow n}  &=& 0.962(14), \nonumber
\eea
which are consistent with the sign of the second-order corrections on $f_1(0)$
reported in previous quenched lattice studies~\cite{{Guadagnoli:2006gj},{Sasaki:2008ha}}, while
our observed tendency of the SU(3) breaking correction 
disagrees predictions of both the latest baryon ChPT result~\cite{Geng:2009ik}
and large $N_c$ analysis~\cite{Flores-Mendieta:1998ii}.

\section{Summary}

\indent\indent
We have presented results of the flavor SU(3) breaking effects on hyperon vector coupling $f_1(0)$
for the $\Xi^0\rightarrow \Sigma^+$ and $\Sigma^{-}\rightarrow n$ decays in (2+1)-flavor QCD using 
domain wall quarks.
We have observed that the second-order correction on $f_1(0)$ is still {\it negative} 
for both decays at much smaller pion mass,
$m_\pi=330$ MeV, than in the previous quenched simulations.
The size of the second-order corrections observed here
is also comparable to what was observed in our DWF calculations of $K_{l3}$ decays~\cite{Boyle:2007qe}.
To extrapolate the value of $f_1(0)$ to the physical point, our simulations at two different sea quark masses
($am_{ud}=0.01$ and 0.02) are now in progress.


\begin{theacknowledgments}

\indent\indent
 We thank C. Jung for help with numerical calculations on the IBM BlueGene/L supercomputer. 
 This work is supported by the JSPS Grant-in-Aids for Scientific Research (C)
 (No.~19540265), Scientific Research on Innovative Areas
 (No.~21105504) and the Large Scale Simulation Program No.09/10-02 (FY2010) of High Energy Accelerator Research Organization (KEK). 
Numerical calculations reported here were carried out at KEK supercomputer system.
\end{theacknowledgments}



\bibliographystyle{aipproc}   


\begin{thebibliography}{99}

\bibitem{Cabibbo:2003cu}
  For a review of hyperon beta decays, see 
  N.~Cabibbo, E.~C.~Swallow and R.~Winston,
  \emph{Ann.\ Rev.\ Nucl.\ Part.\ Sci.}  {\bf 53}, 39 (2003) and references therein.

\bibitem{Gaillard:1984ny}
  J.~M.~Gaillard and G.~Sauvage,
  \emph{Ann.\ Rev.\ Nucl.\ Part.\ Sci.}  {\bf 34}, 351 (1984).

\bibitem{Weinberg:1958ut}
  S.~Weinberg,
  \emph{Phys.\ Rev.\ } {\bf 112}, 1375 (1958).
  
\bibitem{Ademollo:1964sr}
  M.~Ademollo and R.~Gatto,
  \emph{Phys.\ Rev.\ Lett.}  {\bf 13}, 264 (1964).
  
\bibitem{Donoghue:1981uk}
  J.~F.~Donoghue and B.~R.~Holstein,
  \emph{Phys.\ Rev.\  D}{\bf 25}, 206 (1982).

\bibitem{Donoghue:1986th}
  J.~F.~Donoghue, B.~R.~Holstein and S.~W.~Klimt,
  \emph{Phys.\ Rev.\ D} {\bf 35}, 934 (1987).

          
\bibitem{Schlumpf:1994fb}
  F.~Schlumpf,
  \emph{Phys.\ Rev.\ D} {\bf 51}, 2262 (1995).

\bibitem{Flores-Mendieta:1998ii}
  R.~Flores-Mendieta, E.~Jenkins and A.~V.~Manohar,
  \emph{Phys.\ Rev.\ D} {\bf 58}, 094028 (1998).

\bibitem{Villadoro:2006nj}
  G.~Villadoro,
  \emph{Phys.\ Rev.\  D} {\bf 74}, 014018 (2006).

\bibitem{Lacour:2007wm}
  A.~Lacour, B.~Kubis and U.~G.~Meissner,
  \emph{JHEP} {\bf 0710}, 083 (2007).
  
\bibitem{Geng:2009ik}
  L.~S.~Geng, J.~Martin Camalich and M.~J.~Vicente Vacas,
  \emph{Phys.\ Rev.\  D} {\bf 79}, 094022 (2009).


\bibitem{Guadagnoli:2006gj}
  D.~Guadagnoli, V.~Lubicz, M.~Papinutto and S.~Simula,
  \emph{Nucl.\ Phys.\  B} {\bf 761}, 63 (2007).
  
\bibitem{Sasaki:2008ha}
  S.~Sasaki and T.~Yamazaki,
  \emph{Phys.\ Rev.\  D} {\bf 79}, 074508 (2009).
  
\bibitem{Allton:2008pn}
  C.~Allton {\it et al.}  [RBC-UKQCD Collaboration],
  \emph{Phys.\ Rev.\  D} {\bf 78}, 114509 (2008).

\bibitem{Yamazaki:2008py}
  T.~Yamazaki {\it et al.}  [RBC+UKQCD Collaboration],
  \emph{Phys.\ Rev.\ Lett.}  {\bf 100}, 171602 (2008).

\bibitem{Yamazaki:2009zq}
  T.~Yamazaki {\it et al.} [RBC+UKQCD Collaboration],
  \emph{Phys.\ Rev.\  D} {\bf 79}, 114505 (2009).
  
\bibitem{Aoki:2010xg}
  Y.~Aoki {\it et al.} [RBC+UKQCD Collaboration],
  \emph{Phys.\ Rev.\  D} {\bf 82}, 014501 (2010).

\bibitem{Boyle:2007qe}
  P.~A.~Boyle {\it et al.} [RBC+UKQCD Collaboration],
  \emph{Phys.\ Rev.\ Lett.}  {\bf 100}, 141601 (2008).
  
\end{thebibliography}

\IfFileExists{\jobname.bbl}{}
 {\typeout{}
  \typeout{******************************************}
  \typeout{** Please run "bibtex \jobname" to optain}
  \typeout{** the bibliography and then re-run LaTeX}
  \typeout{** twice to fix the references!}
  \typeout{******************************************}
  \typeout{}
 }

\end{document}